\documentclass[12pt,preprint]{aastex}
\usepackage{amsmath}
\usepackage{graphicx}
\usepackage{amsmath,bm}

\begin{document}
\begin{tiny}

\end{tiny}
\title{On the magnetic field of the ultraluminous X-ray pulsar M82 X-2}
\author{Kun Xu$^{1}$and Xiang-Dong Li$^{1,2}$
}
\affil{$^{1}$Department of Astronomy, Nanjing University, Nanjing 210023, China;
lixd@nju.edu.cn}

\affil{$^{2}$Key Laboratory of Modern Astronomy and Astrophysics, Nanjing University,
Ministry of Education, Nanjing 210023, China}

\begin{abstract}
The discovery of the ultraluminous X-ray pulsar M82 X-2 has stimulated lively discussion on the nature of the accreting neutron star. In most of the previous studies the magnetic field of the neutron star was derived from the observed spin-up/down rates based on the standard thin, magnetized accretion disk model. However, under super-Eddington accretion  the inner part of the accretion disk becomes geometrically thick. In this work we consider both radiation feedback from the neutron star and the sub-Keplerian rotation in a thick disk, and calculate the magnetic moment - mass accretion rate relations for the measured rates of spin change. We find that the derived neutron star's dipole magnetic field depends the maximum accretion rate adopted, but is likely $\lesssim 10^{13}$ G. The predicted accretion rate change can be used to test the proposed models by comparison with observations.

\end{abstract}


\section{Introduction}

M82 X-2 (or NuSTAR J095551+6940.8) is an ultraluminous X-ray source (ULX) discovered in the nuclear region of the galaxy M82 \citep{ksl2006,kyh2007,fs2011}. ULXs are very bright X-ray sources with isotropic X-ray luminosities higher than $10^{39}$ ergs$^{-1}$. They have been proposed to be stellar-mass black holes (BHs) in X-ray binaries accreting at a super-Eddington rate, and the ultrahigh luminosities may partly come from geometrical beaming in an accretion funnel \citep{kdw2001, r2007}.
However, the 1.37 s pulsations discovered from M82 X-2 definitely result from the rotation of a magnetized neutron star (NS) rather than a BH \citep{bhw2014}. It's X-ray luminosity  varied between $10^{39}$ and $10^{40}$ ergs$^{-1}$ \citep{b2016}, much higher than the Eddington limit ($L_{\rm E}\sim 10^{38}$ ergs$^{-1}$, corresponding to an accretion rate of $\dot{M}_{\rm E}\sim 10^{18}$ gs$^{-1}$) for a NS with typical mass of $1.4 M_\odot$. Combining with the fact that donor star's mass $>5M_\odot$ \citep{bhw2014}, this indicates that the binary is undergoing rapid mass transfer on a thermal timescale through Roche-lobe overflow \citep{sl2015,flk2015}. More recently, two more NS ULXs were discovered, that is, XMMU J235751.1$-$323725 in NGC7793 P13 with a pulse period of $\sim 0.42$ s  \citep{i2016b,f2016} and ULX-1 in NGC5907 with a pulse period of $\sim 1.1$ s \citep{i2016a}. The peak luminosities for these two sources are $\sim 10^{40}$ ergs$^{-1}$ and $\sim 10^{41}$ ergs$^{-1}$ respectively, also significantly higher than the Eddington luminosity.

NS ULXs provide a unique opportunity to investigate the formation of young NSs in binaries. Obviously their dipole magnetic field strengths play a vital role in determining their nature.
Besides cyclotron line feature measurements \citep[][for reviews] {c2002,cw2012}, there are basically two (indirect) ways to estimate the {\em dipole} magnetic fields of NSs in X-ray binaries. One is from the rates of spin variation. The torque exerted by an accretion disk on a NS is composed by the material torque due to angular momentum transfer from the accreted matter and the magnetic torque originating from the interaction between the NS magnetic field and the disk \citep{gl1979}, both depending on the size of the magnetosphere \citep{bc2017}.  The other assumes that the abrupt change in X-ray luminosity is caused by transition between the accretion and propeller regimes, which occurs when the magnetospheric radius equals the so-called corotation radius. By these ways, the magnetic field of M82 X-2 has been estimated to range from $\lesssim 10^9$ G to $>10^{14}$ G \citep{bhw2014,eac2015,t2015,kl2015,dps2015,dpp2016,ckl2016,tmsp2016,km2016,kl2016,c2016}.

It is noted that the above results critically depend on the inner radius of the accretion disk (or the magnetospheric radius) $r_0$. However, all the previous works except \citet{dpp2016} adopt the standard thin disk model and the traditional Alf\'ven radius to estimate $r_0$, incompatible with the fact that M82 X-2 was accreting at a super-Eddington accretion rate. Recently \citet{dpp2016} considered that the dependence of $r_0$ on the accretion rate $\dot{M}$ changes from $r_0\propto \dot{M}^{-2/7}$ in the gas pressure-dominated regime to $r_0\propto \dot{M}^{-1/7}$ in the radiation pressure-dominated regime when the accretion rate is close to the Eddington limit \citep{g1996}.  We note that the $r_0-\dot{M}$ relation could be more complicated for a (super-)Eddington accreting compact star as indicated by both observational and theoretical studies. For example, \citet{wz2011} and \citet{wzz2014} investigated the disk evolution in several BH and NS X-ray binaries from low to super-Eddington luminosity, and showed that the inner disk radius increases with luminosity when it becomes higher than 0.3 times the Eddington luminosity. A similar conclusion was reached by \citet{cmcmbs2016} in the study of the bright NS system Serpens X-1. From a theoretical point of view, \citet{aghw2005} pointed out that radiation pressure can substantially influence the dynamics and structure of the inner region of the disk for (super-)Eddington accreting sources, which is likely to be dominated by advection rather radiative cooling \citep{ny1994}.

In this paper, we consider the accretion torques in two types (i.e., thin and thick) of disk models to study the spin evolution of M82 X-2.  By comparing the theoretical predictions with observations, we can constrain the NS magnetic field in each model. In particular, we show that the dependence of the spin-up/down rate on luminosity is quite different in these models, and this may offer useful test of the proposed models.

\section{Disk models}

\subsection{Thin disk model}

Our thin disk model is based on the original work of \cite{gl1979} and modified by \citet{w1987, w1995}. In this model the NS magnetic field lines are assumed to thread the accretion disk due to various instabilities and freeze with the disk plasma, so they become twisted and exert a torque on the NS due to differential rotation between the NS and the disk. The inner radius of the disk is given by
\begin{equation}
r_0=\xi r_A,
\end{equation}
where $\xi$ is in the range of $0.5-1$ and $r_{\rm A}$ is the traditional Alf\'ven radius for spherical accretion
\begin{equation}
r_{\rm A}=\left(\frac{\mu^4}{2GM\dot{M}^2}\right)^{1/7},
\end{equation}
where $\mu=BR^3$ is the magnetic dipole moment, $B$ the dipole magnetic field, $M$ the mass, and $R$ the radius of the NS, respectively, and $G$ is the gravity constant. Let the Keplerian angular velocity $\Omega_{\rm K}(r)$ at the radius $r$ equals the angular velocity $\Omega_{\rm s}$ of the NS, one can define the corotation radius $r_{\rm c}$,
\begin{equation}
r_{\rm c}=\left(\frac{GM}{\Omega_{\rm s}^2}\right)^{1/3}.
\end{equation}
If $r_0<r_{\rm c}$, accretion can take place, and the accreted material by the NS transfers angular momentum at a rate of
\begin{equation}
N_0=\dot{M}(GMr_0)^{1/2}.
\end{equation}
Besides this (material) torque, the interaction between the twisted magnetic field lines and the disk exerts an additional torque on the NS. In the inner part of the disk with $r_0<r<r_{\rm c}$ the disk matter rotates more rapidly than the NS, giving a spin-up torque $N_{+}$, while in the outer part with $r>r_{\rm c}$ a spin-down torque $N_{-}$ is yielded. So the total torque on the NS can be written as
\begin{equation}
N=N_0+N_{+}+N_{-}.
\end{equation}
The sum of $N_0$, $N_{+}$ and $N_{-}$ depends on $B$, $\dot{M}$, and $\Omega_{\rm s}$, and can be written in the following form
\begin{equation}
N=N_0f(\omega_s),
\end{equation}
where $f(\omega_{\rm s})$ is a function of  the ``fastness parameter" $\omega_{\rm s}\equiv\Omega_{\rm s}/\Omega_{\rm K}(r_0)$. Since the work of \citet{gl1979}, various forms of the function $f(\omega_s)$ have been worked out under different conditions \citep[e.g.,][]{w1987,w1995,c1997,mp2005,dl2006,kr2007,ds2010}. Here we adopt a simplified form
\begin{equation}
f(\omega_{\rm s})=1-\frac{\omega_{\rm s}}{\omega_{\rm c}},
\end{equation}
where $\omega_{\rm c}$ is the critical fastness parameter when the net torque vanishes, ranging between 0.7 and 0.95 \citep{w1995,lw1996}. There is no essential difference between Eq.~(7) and the more sophisticated ones in the above-mentioned works\footnote{Note that the original form in \cite{gl1979}, which has been used by many authors, was critized by \citet{w1987} to be inconsistent.}. The spin evolution of the NS is then described by
\begin{equation}
I\dot{\Omega}_{\rm s}=N_0\left(1-\frac{\omega_{\rm s}}{\omega_{\rm c}}\right),
\end{equation}
where $I$ is the moment of inertia of the NS. After some transformation, we can rewrite Eq.~(8) to be
\begin{equation}
-\frac{a \dot{P}_{-10}}{\mu_{30}^{2/7}P^2\dot{m}^{6/7}}
=1-\frac{b\mu_{30}^{6/7}}{P\dot{m}^{3/7}},
\end{equation}
where $\dot{P}_{-10}$ is the time derivative of the spin period $P$ in units of $10^{-10}$ ss$^{-1}$, $\dot{m}=\dot{M}/\dot{M}_{\rm cr}$, $\mu_{30}=\mu/10^{30}$ G cm$^3$,
\begin{equation}
a=1.77\times 10^{-18} \xi^{-1/2}I(GM)^{-3/7}\dot{M}_{\rm cr}^{-6/7},
\end{equation}
and
\begin{equation}
b=2.8\times 10^{26}\xi^{3/2}(GM)^{-5/7}\dot{M}_{\rm cr}^{-3/7}{\omega_{\rm c}}^{-1},
\end{equation}
with $\dot{M}_{\rm cr}$ being the maximum accretion rate for a NS, which can be different from and even larger than the traditional Eddington value $\dot{M}_{\rm E}$ (see below).

\subsection{Thick disk model}

For a rapidly accreting NS, radiation inside the disk and from the NS becomes important in determining the dynamics and structure of the inner disk region. In the classic model for super-Eddington accretion a slim/thick inner disk is surrounded by an outer thin disk \citep{acl1988,wft2000,s2011}. The transition radius which connects the inner and outer disks depends on the accretion rate \citep{yn2004,g2012}, and is larger than the corotation radius for M82 X-2 when $L_{\rm X}>10^{39}$ ergs$^{-1}$ \citep{dpp2016}. This means that the traditional thin disk model is  inappropriate for M82 X-2. Two key points should be taken into account for the thick disk model \citep{ywv1997,aghw2005}. First, since the energy inside the inner disk is mainly transported through advection, the rotation of the disk matter becomes sub-Keplerian, that is
\begin{equation}
\Omega(r)=A \Omega_{\rm K}(r),
\end{equation}
with $A<1$. Second, radiation pressure pushes the disk outward, and the inner disk radius is correspondingly modified. In terms of the comoving radiation flux $L_{\rm co}$, one can get the the radiation pressure gradient to be \citep{aghw2005}
\begin{equation}
\frac{{\rm d}P_{\rm rad}}{{\rm d}r}=-\rho \frac{GM}{r^2} \frac{L_{\rm co}}{L_{\rm cr}},
\end{equation}
where $\rho$ is the density in the disk and $L_{\rm cr}$ is the maximum luminosity corresponding to $\dot{M}_{\rm cr}$.
Thus the net gravitational force is $(1-\frac{L_{\rm co}}{L_{\rm cr}})\frac{GM}{r^2}=A^2\frac{GM}{r^2}$, where $A=\sqrt{1-\frac{L_{\rm co}}{L_{\rm cr}}}=\sqrt{1-\dot{m}}$ and $\dot{m}=\dot{M}/\dot{M}_{\rm cr}$.

The sinusoidal pulse profiles observed in both M82 X-2 \citep{bhw2014} and XMMU J235751.1$-$323725 in NGC7793 P13 \citep{f2016} seem to suggest that radiation from the NS is nearly isotropic. However, numerical simulations revealed anisotropic emission for super-Eddington accretion \citep[e.g.,][]{o2007,kmoo2016}. It is unclear how smooth pulse profiles can be produced in such case, and a possible scenario was recently discussed by \citet{msti2016}.
Taking into account possible beamed emission, we can write a more general form as $A=\sqrt{1-\beta\dot{m}}$, where $\beta$ $(<1)$ is the beaming factor. We start from the derivation of the Alf\'ven radius in spherical accretion. The mass density is given by
\begin{equation}
\rho=\frac{\dot{M}}{4 \pi r^2 v_{\rm r}}
\end{equation}
where the radial velocity $v_{\rm r}=A v_{\rm ff}=A \sqrt{2GM/r}$ due to reduced gravitational force, and $v_{\rm ff}$ is the free-fall velocity.
We can estimate the modified Alf\'ven radius $r'_{\rm A}$ from the condition for the balance between the magnetic, radiation, and ram pressures,
\begin{equation}
\frac{B^2}{8 \pi}|_{r'_{\rm A}}=\rho v^2_{\rm r}|_{r'_{\rm A}}=A^2 \rho v^2_{\rm ff}|_{r'_{\rm A}}.
\end{equation}
Thus
\begin{equation}
r'_{\rm A}=A^{-2/7} r_{\rm A}.
\end{equation}
If we adopt the inner disk radius as $r'_0=\xi r'_{\rm A}$, then we have
\begin{equation}
r'_0=A^{-2/7} r_0.
\end{equation}
Note that $r'_0\rightarrow r_0$ when $\bm{\beta\dot{m}\ll 1}$, and increases rather decreases with $\dot{m}$ when $\bm{\beta\dot{m}\rightarrow 1}$, in agreement with the observational results of luminous X-ray binaries \citep{wz2011,wzz2014,cmcmbs2016}.

In the thick disk model, the material torque is changed to be
\begin{equation}
N'_0=A\dot{M}(GMr'_0)^{1/2}=A^{6/7}\dot{M}(GMr_0)^{1/2},
\end{equation}
because of sub-Keplerian rotation, and the fastness parameter is correspondingly,
\begin{equation}
\omega'_{\rm s}=\Omega_{\rm s}/[A\Omega_{\rm K}(r'_0)].
\end{equation}
Similar as in last subsection, we get the equation for the spin evolution
\begin{equation}
I\dot{\Omega}_{\rm s}=N'_0\left(1-\frac{\omega'_{\rm s}}{\omega_{\rm c}}\right),
\end{equation}
or
\begin{equation}
-\frac{a \dot{P}_{-10}}{A^{6/7}\mu_{30}^{2/7}P^2\dot{m}^{6/7}}
=1-\frac{b\mu_{30}^{6/7}}{A^{10/7}P\dot{m}^{3/7}}.
\end{equation}

\section{Results}

We then use Eqs.~(9) and (21) to model the spin variations observed in M82 X-2. Before doing that we first discuss how big $\dot{M}_{\rm cr}$ can be.
 For a given accretion rate $\dot{M}$, the maximum torque exerted on an accreting NS is $\dot{M}(GMr_{\rm c})^{1/2}$ when $r_0$ reaches its maximum value of $r_{\rm c}$. Thus we can estimate the lower limit of the accretion rate from the observed spin-up rate, i.e.,
\begin{equation}
-2\pi I\dot{P}/P^2\leq \dot{M}(GMr_{\rm c})^{1/2},
\end{equation}
or
\begin{equation}
\dot{M}\geq 3.55\times 10^{18}\dot{P}_{-10}P^{-7/3}\,{\rm gs}^{-1}.
\end{equation}
In the above calculation we have adopted $M=1.4M_\sun$ and $I=10^{45}$ gcm$^2$. M82 X-2 experienced a spin-up at a rate of $\dot{P}\simeq -2.7\times 10^{-10}$ ss$^{-1}$ \citep{bhw2014}, so $\dot{M}\geq 4.6\times 10^{18}\,{\rm gs}^{-1}$. The average and largest measured spin-up rates of ULX-1 in NGC 5907 are $-8.1\times 10^{-10}$ ss$^{-1}$ and $-9.6\times 10^{-9}$ ss$^{-1}$, respectively \citep{i2016a}, giving $\dot{M}\geq 1.9\times 10^{19}\,{\rm gs}^{-1}$ and $1.5\times 10^{20}\,{\rm gs}^{-1}$. These values imply intrinsic super-Eddington accretion, which are allowed if the accretion flow is collimated by the NS's strong magnetic field, so that radiation escapes from the sides of the column above the polar regions \citep{bs1976a,bs1976b,mstp2015}. Additionally, strong magnetic fields can reduce the scattering cross section for electrons to be much below the Thompson value \citep{c1971,h1979,p1992}. Recently \citet{kmoo2016} performed a two-dimensional radiation-hydrodynamic simulation of a super-Eddington accretion flow onto a NS through a channeled column, and found that the total luminosity can greatly exceed $L_{\rm E}$ by several orders of magnitude.

Considering the above arguments, we first take $\dot{M}_{\rm cr}=10^{20}$ gs$^{-1}$. Figure 1 presents the calculated $\mu_{30}-\dot{m}$ relations for M82 X-2 in the thin and thick disk models, shown in the upper two and lower four panels, respectively. The four curves correspond to the four measured values of $\dot{P}_{-10}$, i.e., $-0.3$ (the blue line), $\bm{-1.2}$ (the orange line)\footnote{In the \citet{bhw2014} the spin-up/down rates at obsid 7 and 8-9 are not physical values, so we use the average spin-up/down values during that intervals. We thank the the referee for clarification.}, $\bm{-1.87}$ (the green line), and $-2.73$ (the red line) \citep{bhw2014}.
We assume $\xi=0.5$, $\omega_{\rm c}=0.7$ in the left panels, and $\xi=1$, $\omega_{\rm c}=0.95$ in the right panels, respectively. In the middle and lower panels we take $\beta=1$ and $0.1$, respectively.

In the thin disk model (panels a and b),  there can be two solutions of $\mu_{30}$ for a given $\dot{m}$, with the smaller one corresponding to $\omega_{\rm s}\ll \omega_{\rm c}$ and the larger one corresponding to $\omega_{\rm s}\sim \omega_{\rm c}$ \citep[see also][]{dps2015}.
Only the common values of the $\mu_{30}$ for the four curves give a self-consistent estimate of $\mu_{30}$.
In panels a and b they give $\mu_{30}<44$ and $\mu_{30}<11.5$  for thin disk), respectively.

In the thick disk model with $\beta=1$
(panels c and d, with the values of $\xi$ and $\omega_{\rm c}$ same as in panels a and b, respectively),
there is one more branch of solution when $\dot{m}>\sim 0.45$, with $\mu_{30}$ decreasing with $\dot{m}$, so the $\mu_{30}-\dot{m}$ curve becomes closed in the spin-up case. There are actually three branches of common $\mu_{30}$ values for the four curves.
In panel c they are $\mu_{30}<11.3$ for the left and right and left branches, and $\mu_{30}\sim 2\times 10^{-2}$ for the bottom branch;
In panel d they are $\mu_{30}<4.9$ for the left and right branches, and $\mu_{30}\sim 7\times 10^{-3}$ for the bottom branch\footnote{The bottom branch solutions are less favored but not excluded. Almost all young NSs have relatively strong ($>10^{11}$ G) magnetic fields, and low magnetic fields are generally due to extensive accretion episodes, which seem unlikely in high-mass X-ray binaries \citep{bh1991}.}.
When $\beta$ becomes small, the solutions will recover to the the thin disk case. This can be seen that the $\mu_{30}-\dot{m}$ relations in panels e and f become similar to those in the thin disk model, and they give $\mu_{30}<35$ and $\mu_{30}<12.8$, respectively.
The results in Fig.~1 indicate that the inferred dipole field strength of  M82 X-2 can be a few times smaller in the thick disk model than in the thin disk model, with its maximum being $\sim 1\times10^{13}$ G and $\sim 4\times10^{13}$ G, respectively.

We then change the value of $\dot{M}_{\rm cr}$ to explore its influence on the results. When $\dot{M}_{\rm cr}=10^{19}$ gs$^{-1}$, the two models can only account for  the period derivative $\dot{P}=-3.0\times 10^{-11}$, so this choice is disfavored. When $\dot{M}_{\rm cr}=10^{21}$ gs$^{-1}$, we get similar solutions as in the case of $\dot{M}_{\rm cr}=10^{20}$ gs$^{-1}$.
The difference is that larger ranges of the dipole field are obtained. The maximum allowed $\mu_{30}$ is $\bm{\sim 45}$ in the thick disk model with $\bm{\beta=1}$.

Figure 2 shows the predicted mass accretion rates based on the results in Fig.~1 with $\mu_{30}=10$ at the four observational epochs when X-ray pulsations were detected.
The evolutionary trend of $\dot{m}$ with time can be potentially used to test which model can better reproduce the observed ones.
In Fig.~2 the blue dots represent the solutions of the thin disk model and the left branch solutions of the thick disk model with $\bm{\beta=1}$, and the red squares  represent the right branch solutions of the thick disk model with $\bm{\beta=1}$. The light curve in \citet{bhw2014}'s Fig.~1 shows that the count rate of M82 X-2 seems  to slightly increase from obsid 6 to 9, followed by a remarkable decrease towards obsid 11. Note that the blue dots show a constant increase in $\dot{m}$, which is in contradiction with observation of the flux decrease. The red squares reveal a decrease in $\dot{m}$ at obsid 11, but similar trend is also seen from obsid 6 to 9. This could be due to the fact that we have adopted a constant $\beta$ in the thick disk model. In fact $\beta$ might be inversely proportional to $\dot{m}$ \citep{king2009}, so the flux with larger $\dot{m}$ (e.g. at obis 6) could be lower. However, we caution that in the $70\arcsec$-radius region there are two bright sources, i.e., X-1 and X-2, as well as other fainter point sources, and it is not certain whether the observed variation in the X-ray flux was mainly contributed by X-2.

\section{Discussion}

In this work we demonstrate that the observed spin variations in M82 X-2 and the other NS ULXs require that the real accretion rates of the NSs must have been super-Eddington, and in this case the thin disk models adopted in previous works seems not self-consistent. We develop a thick disk model by including both the radiation feedback from the accreting NS and the sub-Keplerian rotating behavior in the inner disk. We show that to account for the observed spin changes, the dipole magnetic field of M82 X-2 is less than a few times $10^{13}$ G, depending on the maximum accretion rate $\dot{M}_{\rm cr}$. This suggests that M82 X-2 is likely a NS with a traditional dipole field $\lesssim 10^{13}$ G.

Assuming that the propeller effect occurs at luminosity of $\sim 10^{40}$ ergs$^{-1}$ in M82 X-2, \cite{tmsp2016} estimated the its magnetic field to be $\sim 10^{14}$ G  by equating the inner disk radius proposed by \citet{gl1979} with the corotation radius.
As we argue before, in the thick disk case, the inner disk radius may deviate from that in the thin disk case.
A comparison of the $\mu-\dot{m}$ relations for the occurrence of the propeller effect in the thin and thick disk models is shown in Fig.~3.
We can see that in the thin disk model $\mu$ always increases with $\dot{m}$ under the condition of $r_0=r_{\rm c}$, so a high $\mu$ is inferred for a high $\dot{m}$. In the thick disk case, $\mu$ becomes smaller for a given $\dot{m}$ with increasing $\beta$. When $\beta=1$, $\mu$ starts to decreases with $\dot{m}$ when $\dot{m}>\sim 0.7$,
thus a very high $\mu$ may not be required. This is consistent with the current view that magnetars are very young NSs (with ages less than a few $10^4$ yr), since ultra-high fields decay by Ohm diffusion and Hall drift on a timescale $<10^{5-6}$ yr \citep[][for a review]{tzw2015}. In high-mass X-ray binaries with a donor star of mass $\sim 5-10\,M_{\odot}$, it usually takes more than $10^7$ yr (i.e., the main-sequence lifetime of the companion star) for the systems to enter the Roche-lobe overflow phase after the birth of the NS \citep{bh1991,sl2015}, so the NS magnetic field may have already decayed into the normal range even if it was born as a magnetar.

\acknowledgements We are grateful to an anonymous referee for very helpful comments. This work was supported by the National Program on Key Research and Development Project (Grant No. 2016YFA0400803, and the Natural Science Foundation of China under grant numbers 11133001 and 11333004.

\newpage

\begin{figure}
\centering
\includegraphics[width=60mm]{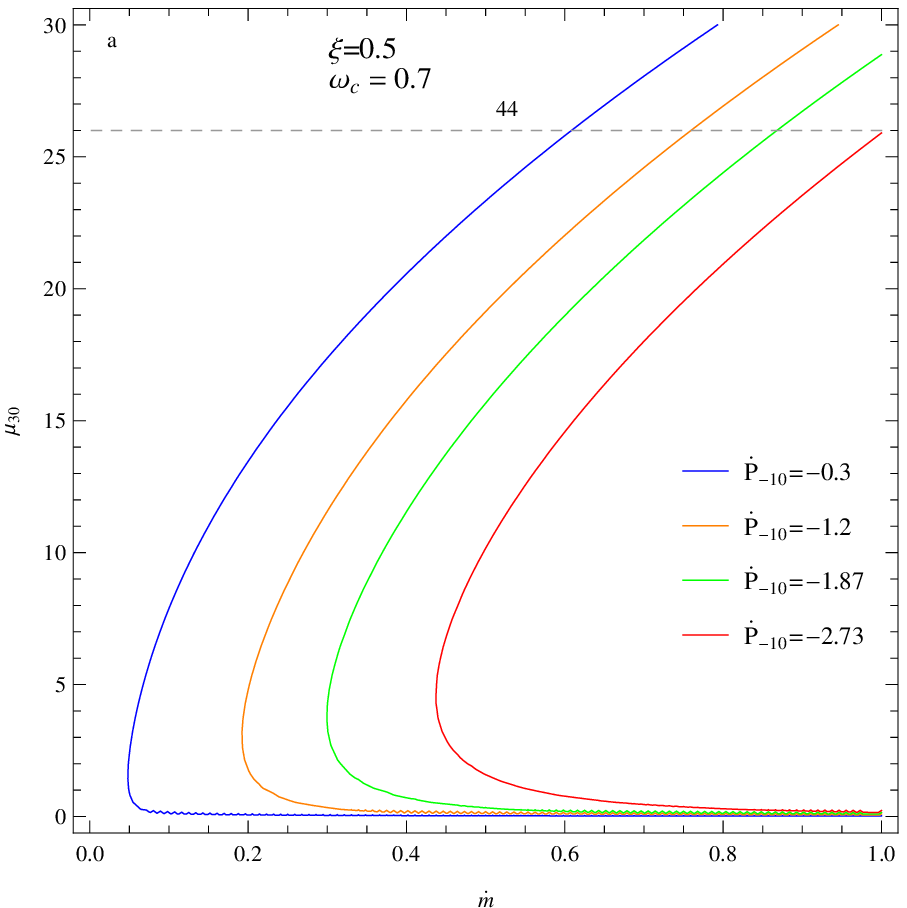}
\includegraphics[width=60mm]{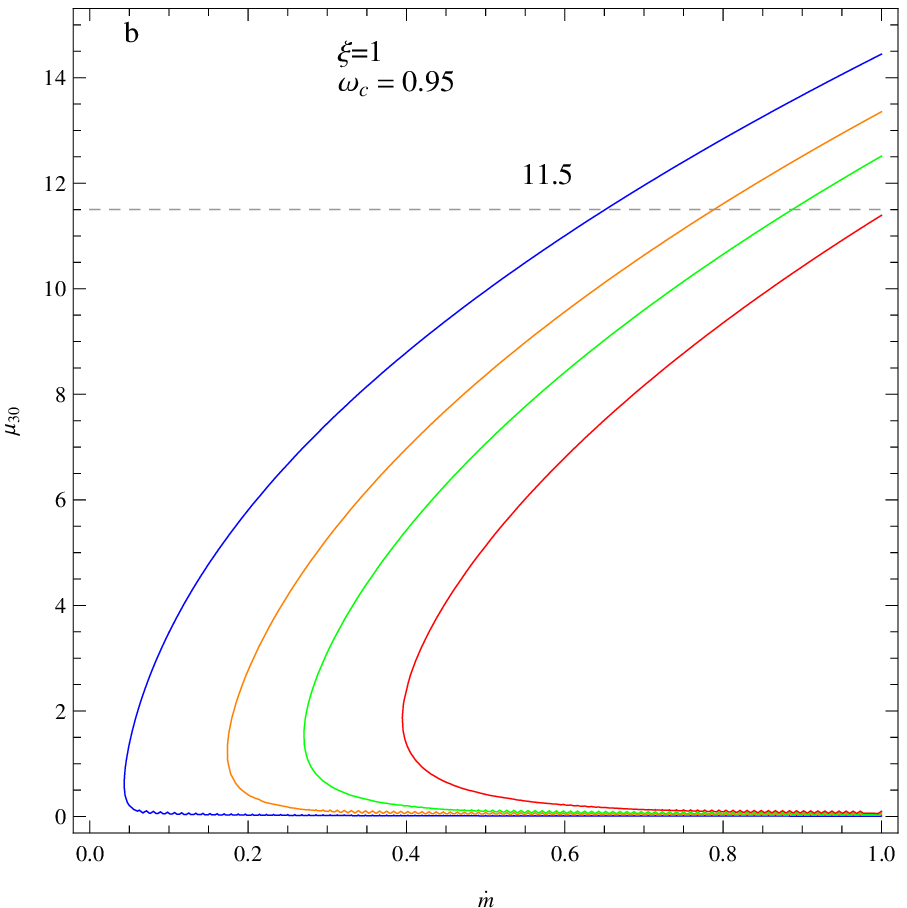}
\includegraphics[width=60mm]{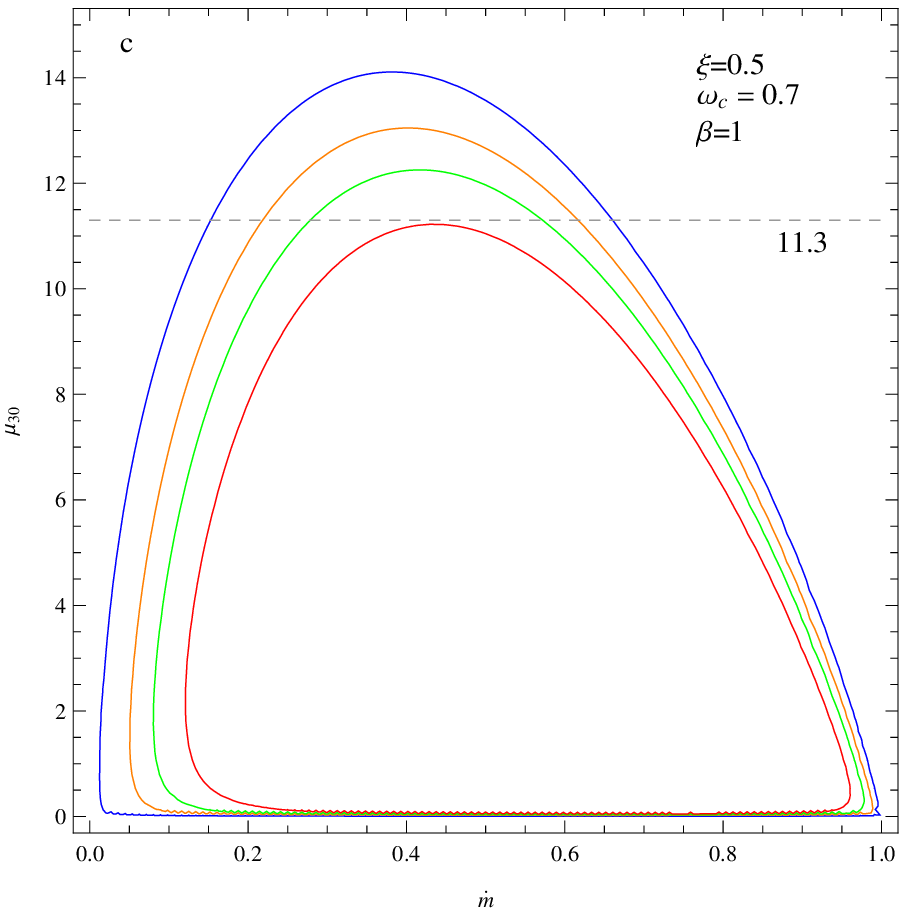}
\includegraphics[width=60mm]{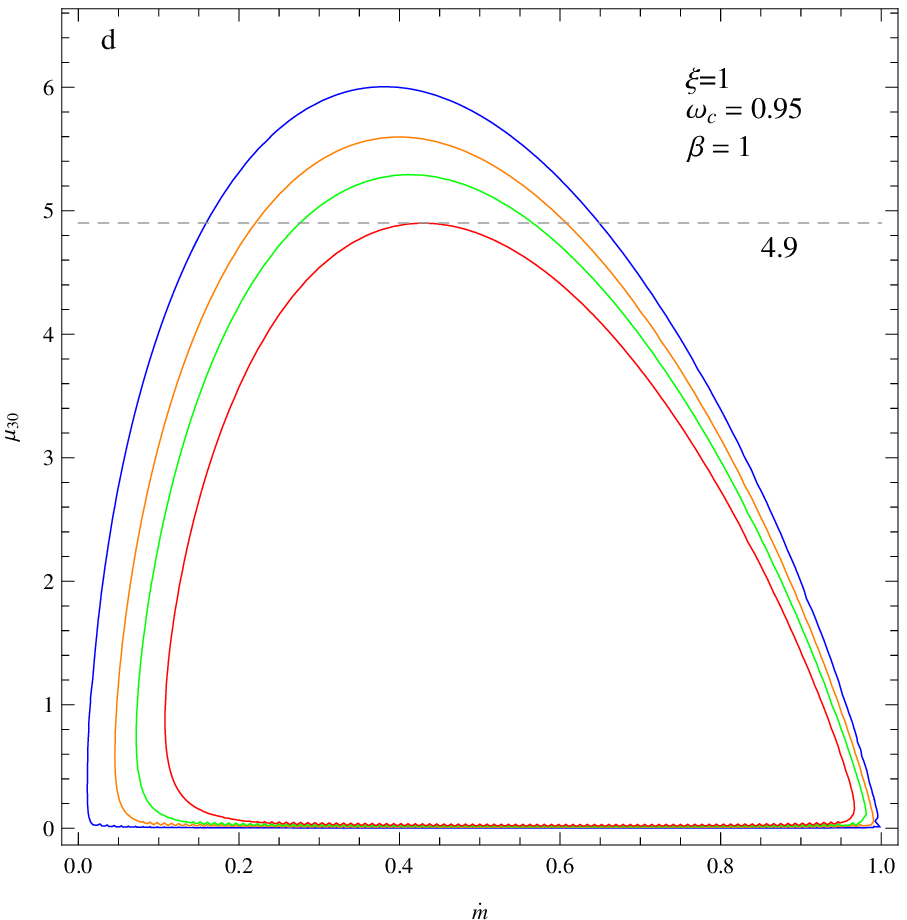}
\includegraphics[width=60mm]{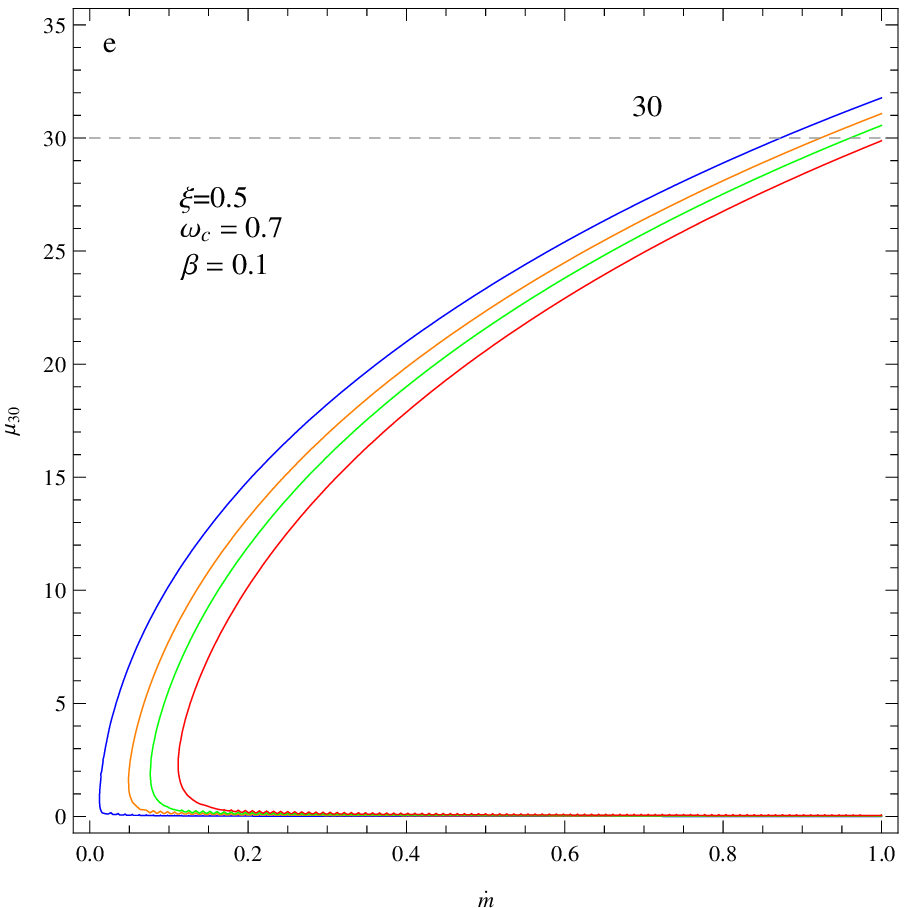}
\includegraphics[width=60mm]{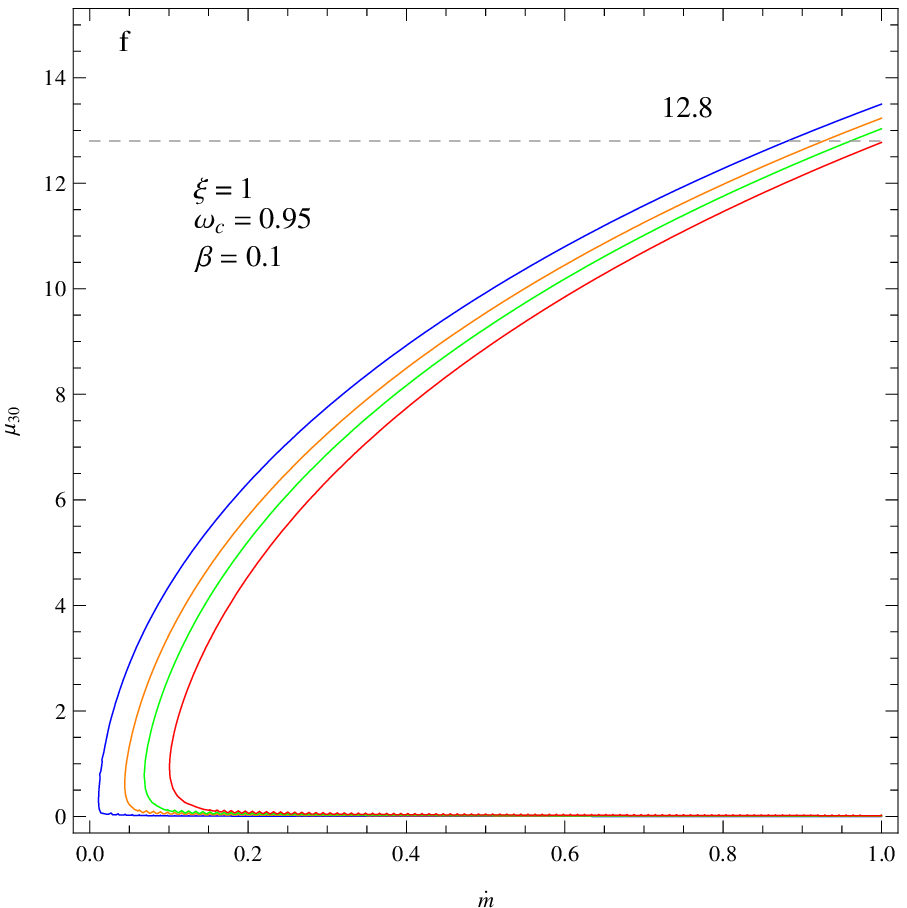}
\caption{The $\mu_{30}-\dot{m}$ relations for M82 X-2 in thin and thick disk models considering all four measured values of the period derivative \citep{bhw2014}, which are $-3.0\times 10^{-11}$ (the blue line), $-1.2\times 10^{-10}$ (the orange line), $-1.87\times 10^{-10}$ (the green line) and $-2.73\times 10^{-10}$ (the red line) (in units of ss$^{-1}$) at obsid 6, 7, 8-9, and 11, respectively. The maximum accretion rate is taken to be $\dot{M}_{\rm cr}=10^{20}$ gs$^{-1}$.
The upper panels are for thin disk model, and the middle and lower panels are for the thick disk model with $\beta=1$ and $\beta=0.1$, respectively. In the left and right panels we take $\xi=0.5$, $\omega_c=0.7$, and $\xi=1$, $\omega_c=0.95$, respectively.
}
%Panel c and d are same as panels a and b, but for the thick disk model, respectively.
\label{fig:pyra}
\end{figure}

\clearpage

\begin{figure}
\centerline{\includegraphics[angle=0,width=1.00\textwidth]{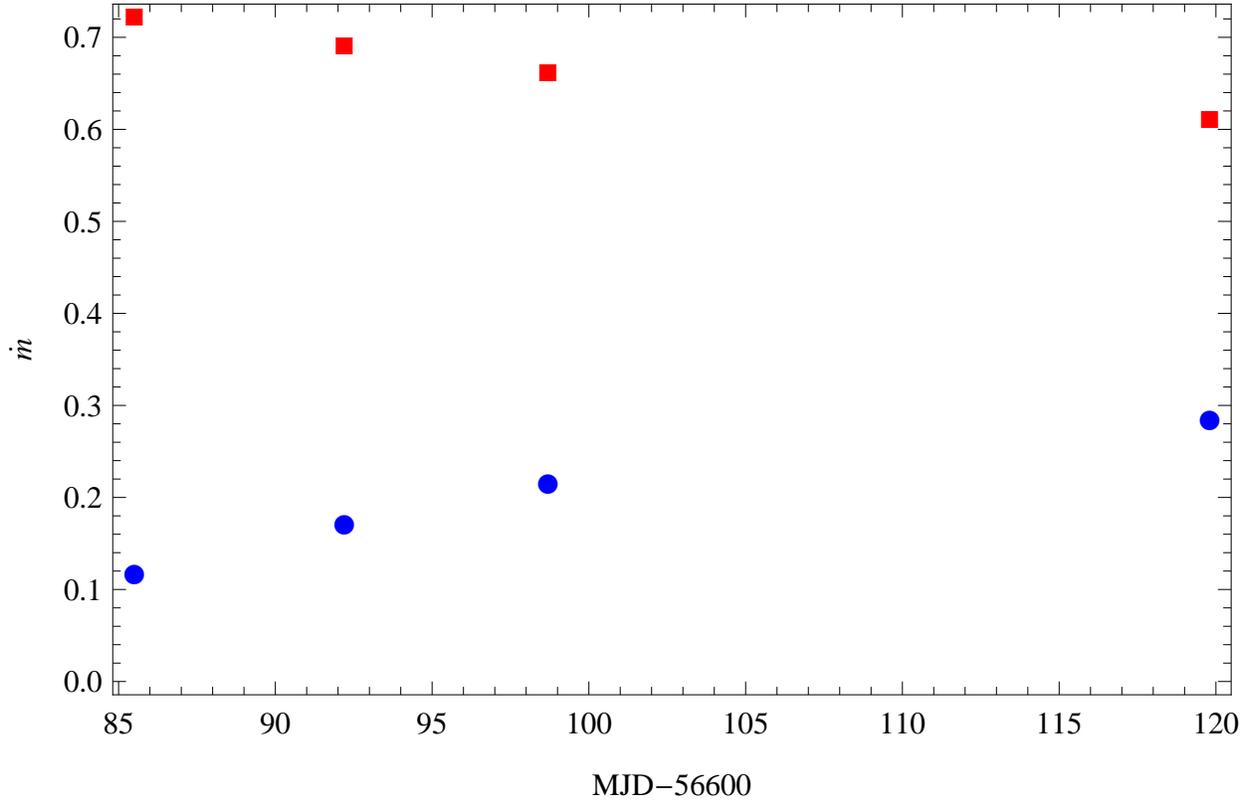}}
\caption{The predicted accretion rate at the four observational epochs with $\mu_{30}=10$. The blue dots represent the solutions of the thin disk model and the left branch solutions of the thick disk model with $\beta=1$, while the red squares represent the right branch solutions in the thick disk model.
}
\end{figure}

\newpage

\begin{figure}
\centerline{\includegraphics[angle=0,width=1.00\textwidth]{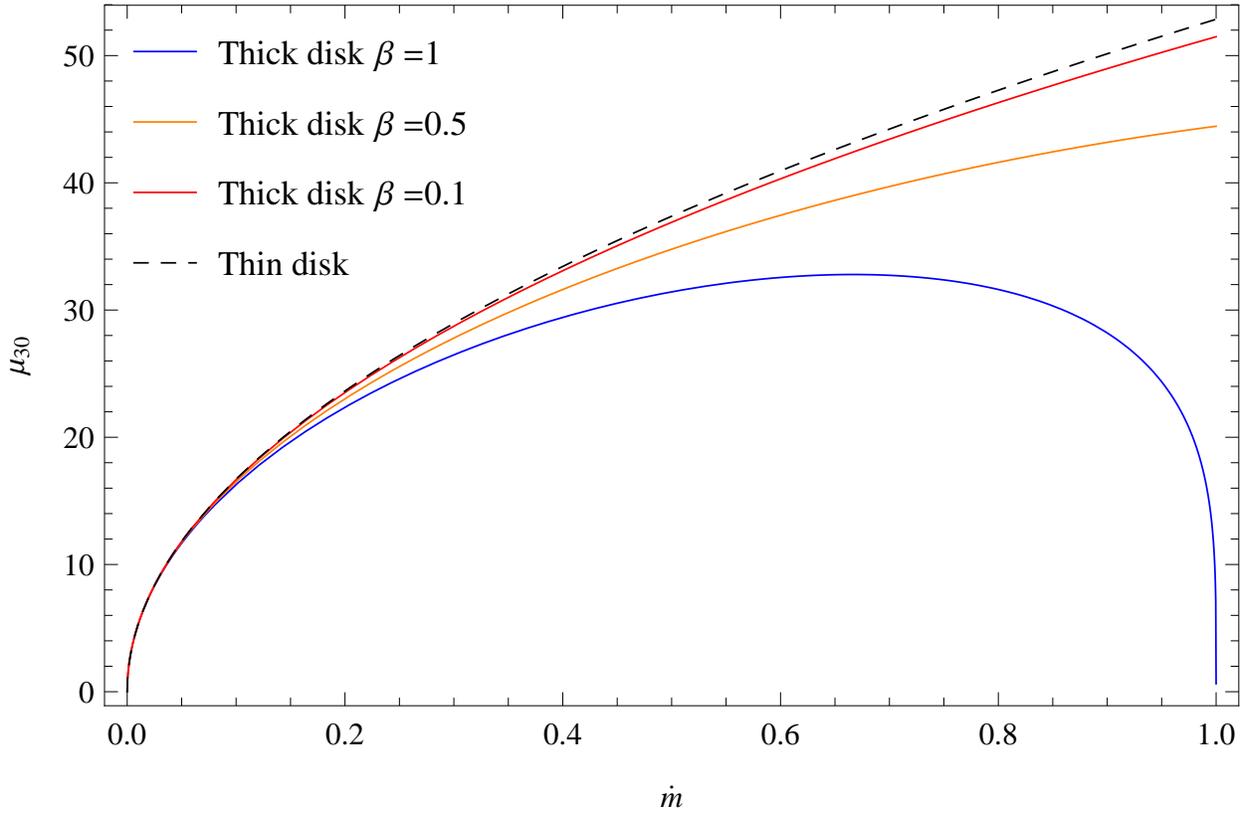}}

\caption{The $\mu_{30}-\dot{m}$ relations with $\xi=0.5$ when the inner disk radius equals the corotation radius, shown with the blue, orange, and red solid line for the thick disk model with $\beta=1$, 0.5, and 0.1, and the dashed line for the thin disk model, respectively.}
\end{figure}

\label{lastpage}
\end{document}